# Analytical photoresponses of gated nanowire photoconductors

Yinchu Shen[1], Jiajing He[1,2*], Yang Xu[3], Kaiyou Wang[4*], Yaping Dan[1*]

[1]University of Michigan – Shanghai Jiao Tong University Joint Institute, Shanghai Jiao Tong University, Shanghai 200240, China

[2]Aerospace Laser Technology and System Department, Shanghai Institute of Optics and Fine Mechanics, Chinese Academy of Sciences, Shanghai 201800, China

[3]School of Micro-Nano ElectronicsZhejiang UniversityHangzhou 310027, China

[4]State Key Laboratory for Superlattices and Microstructures, Institute of Semiconductors, Chinese Academy of Sciences, Beijing 100083, China

Email: yaping.dan@sjtu.edu.cn, kywang@semi.ac.cn, jiajinghe@siom.ac.cn

Abstract

Low-dimensional photoconductors have extraordinarily high photoresponse and gain, which can be modulated by gate voltages as shown in literature. However, the physics of gate modulation remains elusive. In this work, we investigated the physics of gate modulation in silicon nanowire photoconductors with the analytical photoresponse equations. It was found that the impact of gate voltage varies vastly for nanowires with different size. For the wide nanowires that cannot be pinched off by high gate voltage, we found that the photoresponses are enhanced by at least one order of magnitude due to the gate-induced electric passivation. For narrow nanowires that starts with a pinched-off channel, the gate voltage has no electric passivation effect but increases the potential barrier between source and drain, resulting in a decrease in dark and photo current. For the nanowires with an intermediate size, the channel is continuous but can be pinched off by a high gate voltage. The photoresponsivity and photodetectivity is maximized during the transition from the continuous channel to the pinched-off one. This work provides important insights on how to design high-performance photoconductors.

Low-dimensional photoconductors have been extensively investigated in the past several decades on thin films,[1,,2] nanowires,[3,4] quantum dots[5,6] and more recently two-dimensional semiconductors.[7,8] The persistent research interests were mainly driven by the extraordinarily high photogain observed in these devices (up to $10^{10}$).[9,10] According to the classical photogain theory,[11] high gain and high bandwidth can be achieved in photoconductors as long as the device is made short enough, which however has never been realized.[12] In fact, the classical theory was derived on two misplaced assumptions as shown in our recent publication.[13] The first assumption is that the photogenerated excess carriers are uniform in spatial distribution on condition of uniform doping and illumination, which becomes invalid in the presence of metal-semiconductor boundary confinement. The second assumption is that the excess electrons and holes equally contribute to the photocurrent, which intuitively should be true because the absorption of photons will create an equal number of electrons and holes. However, devices in practice often have surfaces or interfaces where depletion regions and/or trap states appear. The depletion regions and/or trap states will localize excess minority carriers, leaving the majority counterparts to dominantly contribute to the photocurrent.

After correcting these two assumptions, we derived a theoretical photogain equation which, however, like the classical theory, is an implicit function of light intensity and device parameters (physical size, doping concentration, etc).[13] As a result, it cannot be used for fitting experimental data or guiding the device design. To search for an explicit photogain theory, we analyzed the dark and photo current that are dependent on nanowire size, finding that all nanowires have a surface depletion region which plays an important role in photogain. Photo Hall measurements showed that the built-in electric field in surface depletion separates photogenerated electrons and holes, pumping excess majority carriers into the conduction channel.[3] The separation of excess electrons and holes will create a forward photovoltage that narrows down the surface depletion region and widens the conduction channel. In return, a large photocurrent and photogain are induced.[14]

Interestingly, after the channel is widened, the actual increase of excess carrier concentration in the conduction channel is negligible (also true for bulk device).[14] This means that a high-gain photoconductor can be modeled as a resistor with a channel that has a width controlled by the floating surface junctions. Following this model, we established an explicit photogain theory that can fit well the experimental data, from which we extracted the doping concentration, carrier mobility, minority carrier recombination lifetime and other parameters.[14] The extracted doping concentration and carrier mobility were verified with Hall effect measurements.[15] The minority carrier recombination lifetime was validated independently by near-field scanning photocurrent microscopy.[16]

Gate-modulated photoconductors exhibit interesting photoresponses, which were explored extensively in recent years.[17,18,19] In this work, we investigated the physics of gate modulation in silicon nanowire photoconductors with the analytical photoresponse equations. For the wide nanowires that cannot be pinched off by gate voltage, we found that the photoresponses are enhanced by at least one order of magnitude as the gate voltage increases, reaching a saturation when the surface depletion region reaches its maximum. The fitting of the analytical photoresponses with the experimental data shows that the photoresponse enhanced by the gate is mainly caused by the electric passivation effect instead of the widening of the surface depletion region. For narrow nanowires that start with a pinched-off channel, the photoresponses decrease as the gate voltage increases. The surface depletion region is forced by the nanowire width. A higher gate voltage does not increase surface depletion region and therefore will not induce a stronger electric passivation effect. Instead, a higher gate voltage will only increase the potential barrier between source and drain, resulting in smaller dark and photocurrent. For the nanowires with an intermediate size, the channel is continuous but can be pinched off by a high gate voltage. Accordingly, the photoresponses start with a pattern similar to the channel-continuous nanowire and then transit to one like the pinched-off nanowire as the gate voltage increases.

**Results and Discussion**

The silicon nanowires were fabricated by patterning the 220 nm thick device layer of a silicon-on-insulator (SOI) wafer which was pre-doped with boron at a concentration of ~$8.6 \times 10^{17}\ cm^{-3}$ by ion implantation (See experimental section for details). These nanowires are 30 μm long with different widths ranging from 50 nm to 230 nm. Each of these nanowires is connected with two large Si pads (formed along with nanowires) to form negligible Ohmic contacts with Al electrodes. Figure 1a shows the optical microscopic (OM) image of the device. A 50 nm thick $HfO_2$ was formed around the silicon nanowires as the gate dielectric, followed by thermal evaporation of Indium Tin

Oxide (ITO) as gate electrodes. The transparent ITO electrode can act as both metal gate and transmission window for light illumination (Transmittance for ITO gate and HfO$_2$ insulator layer shown in Supplemental Information Section 1). Figure 1b shows the I-V curves for the as-fabricated nanowires which transit from a linear correlation into a nonlinear one when the nanowires are narrower than ~ 80 nm. This phenomenon indicates the existence of surface depletion regions, which was widely observed in our previous experiments.[3,14-16]

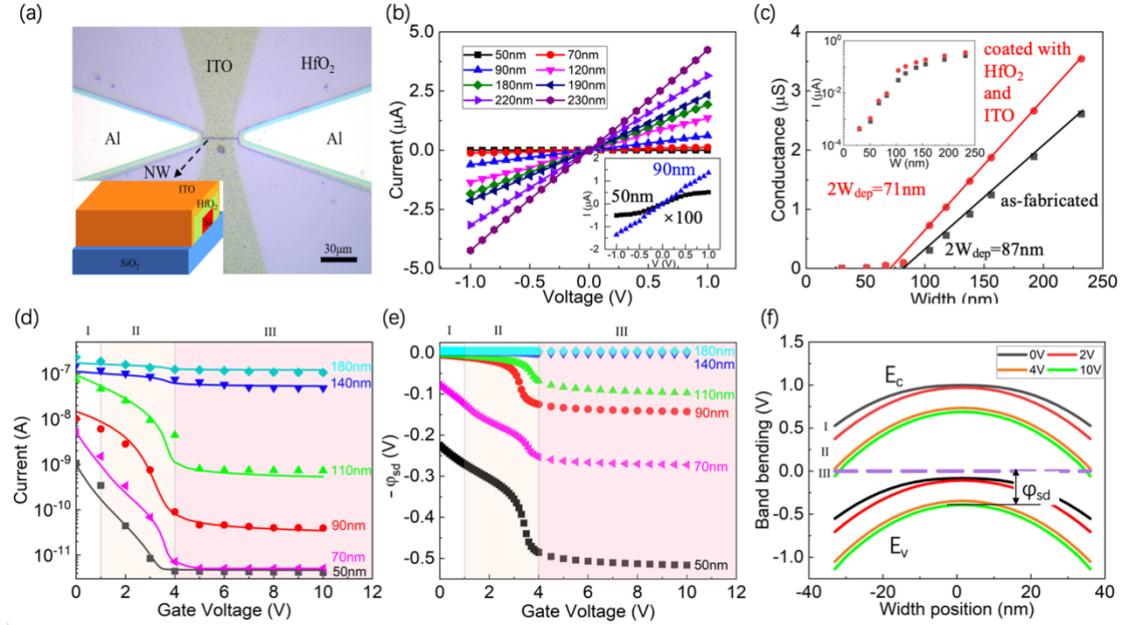

Figure 1. (a) Optical microscopic image of a silicon nanowire photoconductor with aluminum electrodes, HfO$_2$ dielectric and ITO gate. Inset: device schematic. (b) I-V characteristics of silicon nanowires with a width ranging from 50 nm to 230 nm. (c) Width-dependent conductance of silicon nanowires before and after coated with HfO$_2$ and ITO. (d) Gate transfer characteristics of nanowire field effect transistors. Symbols are experimental data and solid lines are simulation results. (e) Simulated negative potential barrier as a function of gate voltage for nanowires with different width. (f) Simulated potential profile along the nanowire width for the domain I, II and III in (d) and (e).

The current for the nanowire with a continuous channel can be calculated following eq.(1).

$$I = \mu N_A q E (H - 2W_{dep})(W - 2W_{dep}) \qquad (1)$$

, in which $\mu$ is the mobility of holes, $N_A$ the doping concentration, $q$ the unit charge, $H$ and $W$ the physical height and width of the nanowire and $W_{dep}$ the depletion region width. As the nanowire width W narrows down, the current at a fixed bias linearly decreases, as shown in Fig.1c. Extending this linear correlation, we find that the intercept equals to $2W_{dep}$ according to eq.(1). Clearly, we have $W_{dep} \simeq 43.5$ nm for the as-fabricated nanowires. After the nanowires were coated with $HfO_2$ and Indium Tin Oxide (ITO), the depletion region width decreases to $W_{dep} \simeq 35.5$ nm due to the large work function (compared to our Si nanowire) of ITO and surface passivation effect of $HfO_2$. When the nanowire is narrower than $2W_{dep}$, the nanowire channel is pinched off and a potential barrier is created between source and drain. As a result, the current exponentially drops as the nanowire width decreases (see the inset of Fig.1c).

Fig.1d depicts the gate dependent source-drain current for nanowires with different widths. To clearly explain the gate dependent behavior, we divide the gate dependent current into three domains (I, II and II) and also simulate the gate dependent potential in Fig.1e in addition to the potential profile in Fig.1f. First, let us take a look at the current-gate dependency among different nanowires. It is clear that the current is mostly determined by the potential barrier between source and drain because the gate dependent current in Fig.1d and potential in Fig.1e follow almost the same pattern except for those very narrow nanowires. For wide nanowires (140 nm and 180 nm), the channel is too wide to pinch off. The potential barrier $\varphi_{sd}$ is zero and the current only slightly (sub-linearly) decreases as the gate voltage increases because the surface depletion region is expanding to its maximum ($W_{dep}^{max}$) of ~ 50 nm in strong inversion from the initial width $W_{dep}^{0}$ ~ 35 nm at $V_g = 0$ V. For the very narrow nanowires, the potential barrier $\varphi_{sd}$ is large and the current of majority holes across the barrier becomes too small. Instead, the current is dominated by the leakage current between channel and source/drain. This is why the current is the same for the 50 nm and 70 nm wide nanowires when $V_g > 5V$ (Fig.1d).

Second, we focus on the three domains of each nanowire in Fig.1d and 1e. In the domain I, the surface depletion region is small. The increase of gate voltage will enlarge the surface potential but not the potential barrier $\varphi_{sd}$ (see black and red line in Fig.1f) unless the nanowire is pinched off because of narrow width (≤ 70nm). As the gate voltage increases to the domain II, the surface depletion region transits from depletion to weak inversion ($E_i$ across $E_F$ near the surfaces). The surface charge concentration is low and charge carriers in the channel center have to get involved in response to the gate voltage increase. As a result, the potential barrier $\varphi_{sd}$ is highly dependent on the gate voltage (from red line to orange line in Fig.1f) and the current quickly drops in this domain (Fig.1d). In the domain III, the surfaces are in strong inversion and the surface electrons respond to the increase of gate voltage. As a result, the potential barrier φ$_{sd}$ is weakly dependent on gate voltage (see the orange and green lines in Fig.1f) according to the principle of classical metal-oxide-semiconductor (MOS) capacitors. This results in the saturation of current in the domain III.

Now, let us discuss the gate dependent photoresponses of these nanowires. In Fig.1d, the nanowires can be roughly divided into three categories. The first category is those nanowires wider than $2W_{dep}^{max}$ that can never be pinched off by the gate voltage (180 nm and 140 nm wide). The second category is those narrower than $2W_{dep}^{0}$ that start with a pinched-off channel at V$_g$ = 0 V (70 nm and 50 nm wide). After we have a better understanding of these two extreme cases, we are ready to discuss the third category between the first and second one in which the nanowires will transit from a continuous channel to a pinched-off one as the voltage increases (110 nm and 90 nm wide).

**Nanowires with a width $W > 2W_{dep}^{max}$.** These nanowires are wide enough to always maintain a continuous channel even at high gate voltage. In this case, the surface depletion region acts like a floating Schottky junction. The nanowire photoconductor can be modeled as a photoresistor that has a channel width controlled by the floating Schottky junction according to our previous publication.[14] The Schottky junction is modulated by both the gate voltage and light illumination. Fig.2a shows the source-drain current dependent on gate voltage under light illumination. As the voltage ramps up, the depletion region increases to its maximum, minimizing the conduction channel width and also the current. Light illumination narrows down the depletion region because the separation of excited electron-hole pairs by the built-in electric field creates a forward photovoltage. As a result, the conduction channel is broadened and a photocurrent (photoconductance) is created in the conduction channel. In such a scenario, we expect that the light illumination is more like a photogate, resulting in a right shift of the current along V$_g$. Surprisingly, the light illumination has some effect of lifting up the current at high gate voltage.

To further explore this phenomenon, we plot in Fig.2b the photocurrent *vs* gate voltage under the illumination of different light intensity calculated from Fig.2a. As the gate voltage increases from 0 to 10 V, the photocurrent is increased by at least one order of magnitude although the depletion width is increased by only a factor of 2 (estimated from the dark current decrease in Fig.2a). The increase in photocurrent by gate voltage is mainly driven by the electric passivation effect (see more discussions below). The photocurrent eventually saturates at high gate voltage simply because the electric passivation effect peaks as the depletion region width reaches its maximum.

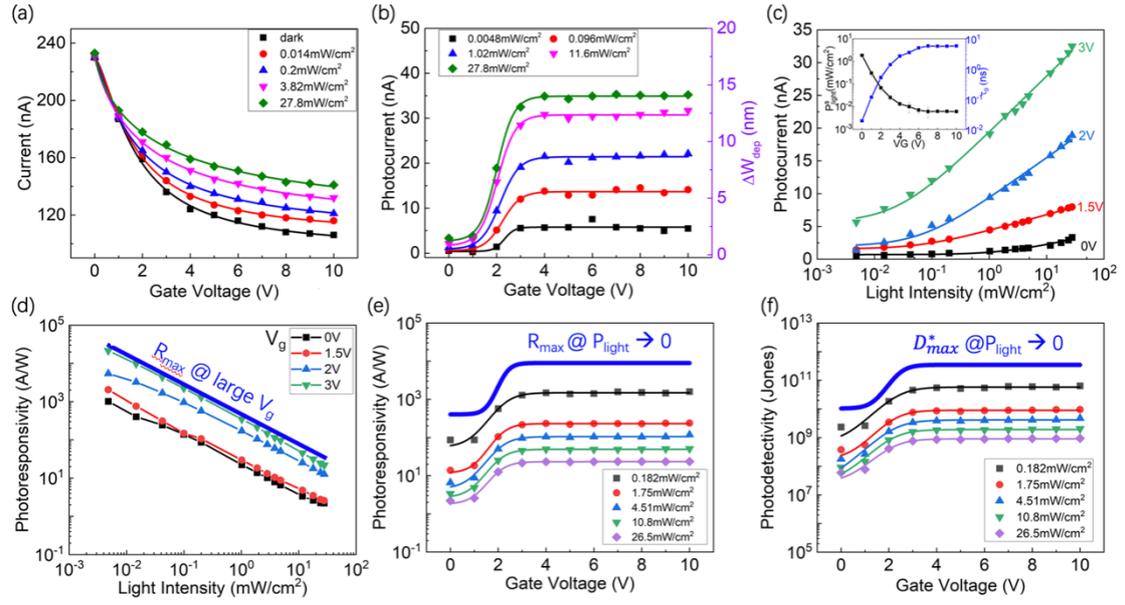

Figure 2. Gate transfer optoelectronic characteristics for the nanowire device 180nm wide at a fixed bias of 0.1V. (a) Current *vs* gate voltage under different light intensity. (b) Photocurrent *vs* gate voltage under different light intensity. (c) Photocurrent *vs* light intensity under different gate voltage. Inset: Extracted critical light intensity $P_{light}^s$ and minority carrier lifetime $\tau_0$. (d) Photoresponsivity *vs* Light intensity under different gate voltage, which reaches the maximum at large $V_g$ (blue line). (e) Photoresponsivity *vs* gate voltage under different light intensity which reaches the maximum as the light intensity approaches zero (blue line). (f) Photodetectivity *vs* gate voltage under different light intensity which reaches the maximum as the light intensity approaches zero (blue line).

To quantitatively show the electric passivation effect, we plot the photocurrent vs light intensity at different gate voltages in Fig.2c. According to our previous publication,[14] the photocurrent of a channel-continuous nanowire is correlated with the light illumination intensity following eq.(2). Fitting the experimental data with eq.(2) allows us to extract the critical light intensity $P_{light}^s$ from which we further calculate the effective minority recombination lifetime $\tau_0$ that is gate-dependent as shown in the inset of Fig.2c (see SI Section2). The lifetime $\tau_0$ increases by more than 2 orders of magnitude due to the electric passivation effect as the gate voltage sweeps from 0 to 10V.

$$I_{ph} = I_{th} \ln\left(\frac{P_{light}}{P_{light}^S} + 1\right) \tag{2}$$

, where $I_{th} = \frac{\eta k_B T}{2qV_{bi0}} I_{ph}^S$ is defined as threshold photocurrent, while $P_{light}^S$ is defined as critical light intensity. $\eta$ is the ideality factor for the surface depletion as a floating Schottky junction, $k_B$ is the Boltzmann constant, T the absolute temperature, q the unit charge and $V_{bi0}$ the built-in potential of the surface depletion region. $I_{ph}^S = \mu pqEW_{dep}\left|\frac{dA_{ch}}{dW_{dep}}\right|$ is defined as the critical photocurrent in which $\left|\frac{dA_{ch}}{dW_{dep}}\right|$ is defined as the absolute derivative of the conducting channel area respective to the depletion region width. Note that the expression of the critical photocurrent is valid only when the source-drain bias is relatively small. When the source-drain bias is large enough, the photocurrent at a fixed light intensity will eventually saturate due to the metal-semiconductor boundary confinement. The critical light intensity is given by $P_{light}^S = \frac{\hbar\omega J_s H}{\alpha W_{dep} q}$ with $\hbar\omega$ being the photon energy, $J_s$ the leakage current density of the surface depletion region as a floating junction, $\alpha$ the photon absorption ratio (see SI Section1), $W_{dep}$ the depletion region width. Additionally, the leakage current density $J_s$ is written as $J_s = \frac{qn_i W_{dep}}{2\tau_0}$ with $n_i$ being the electron concentration in intrinsic silicon and $\tau_0$ is the effective minority recombination in doped silicon.

Fig.2d exhibits the light-intensity-dependent photoresponsivity at different gate voltages. The photoresponsivity R can be written as eq.(3) according to the definition and eq.(2). When the light intensity $P_{light}$ reduces to zero, the term after $R_{max}$ on the right side of eq.(3) increases towards 1 and therefore the photoresponsivity $R$ approaches to its maximum value $R_{max}$. The photoresponsivity $R$ becomes larger at a higher gate voltage but will reach a limit (blue solid line) as the photocurrent will saturate at high gate voltage as shown in Fig.2b. We can also take a look at the photoresponsivity $R$ as a function of gate voltage at different light intensities, shown in Fig.2e. The trend is similar to the photocurrent dependency on gate with a saturation at high gate voltage. The photoresponsivity increases to a limit (solid blue line) as the light intensity decreases to zero. Note that, according to eq.(3), this limit is dependent on the nanowire physical parameters including minority recombination lifetime $\tau_0$, doping concentration, depletion region width and nanowire size.

$$R = \frac{I_{ph}}{P_{light}WL} = R_{max}\frac{P^S_{light}}{P_{light}}\ln\left(\frac{P_{light}}{P^S_{light}} + 1\right) \qquad (3)$$

, in which $R_{max} = \frac{I_{th}}{P^S_{light}WL} = \frac{2\alpha q W_{dep}}{\hbar\omega A_c}\frac{\tau_c}{\tau_t}\left|\frac{dA_{ch}}{dW_{dep}}\right|$ with $\alpha$ being the light absorption ratio by the nanowire, $\hbar\omega$ the photon energy, $A_c$ the nanowire cross-sectional area, $\tau_c$ transient decay time constant and $\tau_t$ the majority carrier transit time. More specifically, the transient decay time constant $\tau_c = \frac{\eta k_B T N_A}{2q n_i V_{bi0}}\tau_0$ can be found from our previous work where $\eta$ is the ideality factor of the surface depletion region as a floating Schottky junction, $k_B$ the Boltzmann constant, T the absolute temperature, $N_A$ the doping concentration, $n_i$ the intrinsic carrier concentration and $V_{bi0}$ the built-in voltage of the surface depletion region in darkness. The majority carrier transit time is expressed as $\tau_t = \frac{L}{\mu E}$ where L is the nanowire length and E is the electric field intensity under the source-drain bias.

Photoresponsivity only represents the performance of photodetectors in one aspect. Photodetectivity $D^*$ is a more comprehensive parameter to gauge the performance of a photodetector that can balance the photoresponsivity and noises. The power spectrum of thermal noise is given as $\overline{I_n^2} = \frac{4k_B T}{r}$ with r being the nanowire resistance in darkness, $k_B$ the Boltzmann constant and T the absolute temperature.[20,21] The analytical expression of photodetectivity is given in *eq.(4)* for the channel-continuous nanowires. Similar to *eq.(3)*, the photodetectivity in *eq.(4)* will approach to its maximum $D^*_{max}$ as the light intensity reduces to zero (blue solid line in Fig.2f). $D^*_{max}$ is linearly proportional to $R_{max}$ and the square root of resistance $\sqrt{r}$ in darkness. The sweep of incremental $V_g$ reduces the dark current by a factor of ~2 (r increases by 2 times) for the 180 nm wide nanowire (Fig.2a) but increases the responsivity by at least one order of magnitude due to the electric passivation effect (Fig.2e). For this reason, the photodetectivity is dominated by photoresponsivity, therefore exhibiting a dependence on gate voltage similar to the photoresponsivity, as shown in Fig.2e and f. As the gate voltage ramps up, $D^*_{max}$ increases from $\sim 10^{10}$ Jones to $\sim 5 \times 10^{11}$ Jones for this 180 nm wide nanowire. Note that the detectivity will increases proportionally as the source-drain bias increases unless the drift length ($L = \mu E \tau$) is comparable to the nanowire length (30μm), for which the source-drain bias has to reach as high as $\sim$ 30V because the effective minority recombination lifetime in the nanowires is as short as nanoseconds.

$$D^* = \frac{R\sqrt{A}}{S_n} = \frac{R\sqrt{WL}}{\sqrt{\frac{4k_BT}{r}}} = D^*_{max} \frac{P^S_{light}}{P_{light}} \ln\left(\frac{P_{light}}{P^S_{light}} + 1\right) \quad (4)$$

, in which $D^*_{max} = \frac{2\alpha q W_{dep}}{\hbar\omega A_c} \frac{\tau_c}{\tau_t} \sqrt{\frac{WLr}{4k_BT}} \left|\frac{dA_{ch}}{dW_{dep}}\right|$.

**Nanowires with a width $W < 2W^0_{dep}$.** Let us then focus on the nanowires that are pinched-off at $V_g = 0$ V (see black squares in Fig.3a). The pinched-off nanowire will be switched back to the channel continuous mode if the negative gate voltage is applied (See SI Fig.S3). For the pinched-off nanowire, there is a the potential barrier φ$_{sd}$ between source and drain. Light illumination will reduce φ$_{sd}$ and create a photocurrent $I_{ph}$ (Fig.3a). The fact that the current is dominated by the potential barrier $\varphi_{sd}$ allows us to find the following correlation of photocurrent and the variation of $\varphi_{sd}$ as shown in eq.(5).[14]

$$\frac{I_{dark} + I_{ph}}{I_{dark}} = \exp\left(\frac{q\Delta\varphi_{sd}}{kT}\right) \quad (5)$$

, in which k is the Boltzmann constant, T the absolute temperature and I$_{dark}$ the dark current.

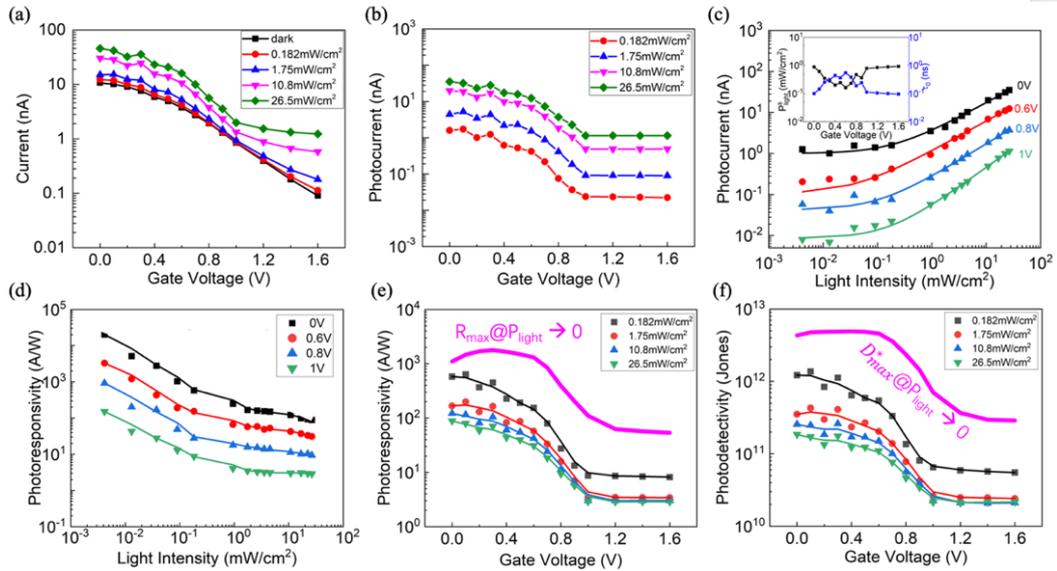

Figure 3. Gate transfer optoelectronic characteristics of the nanowire device 70 wide at a fixed bias of 1V. (a) Current *vs* gate voltage under different light intensity. (b) Photocurrent *vs* gate voltage under different light intensity. (c) Photoresponsivity *vs* light intensity under different gate voltage. (e) Photoresponsivity *vs* gate voltage under different light intensity, which will maximumize (pink lin) as light intensity approaching zero. (f) Photodetectivity *vs* gate voltage under different light intensity, which will maximize (pink line) when light intensity approaching zero.

The photocurrents dependent on gate voltage at different light intensities are plotted in Fig.3b. The photocurrent follows a similar pattern of dark current dependent on gate voltage. As mentioned previously, when the negative gate voltage is applied, the nanowire will be switched to the channel-continous mode. In this case, the photocurrent will follow a patten in Fig.2b, that's, the photocurrent declines as the gate voltage moves towards more negative (see SI Fig.S3). What's weird in Fig.3b is that the photocurrent levels off at $V_g > 1V$. To better understand this level-off phenomenon, we need to find how the minority recombination lifetime changes on the gate voltage, since previous data in Fig. 2c show that the minority recombination lifetime is a dominant factor for photocurrent. For this reason, we plot in Fig.3c the photocurrent dependent on light intensity at different gate voltages. According to our previous work, the dependence of photocurrent on light intensity for channel-pinched-off nanowires can be written as eq.(6).[14] Fitting the experimental data with eq.(6) allows us to extract the critical light intensity $P_{light}^s$ (black line) and then the minority recombination lifetime $\tau_0$ (line line) as shown in the inset of Fig.3c. See SI Section3 for fitting results. Overall, the pinched-off nanowire has a relatively poor electric surface passivation as the minority recombination lifetime $\tau_0$ is within a range of 0.1ns to 1ns, 1-2 orders of magnitude lower than the channel-continuous nanowire at high gate voltage (see the inset of Fig. 2c). A closer look indicates that $\tau_0$ first increases and then decreases and finally levels off at $V_g > 1V$. Accordingly, the electric passivation effect and the resultant photocurrent will level off at high gate voltage in Fig.3b.

$$I_{ph} = I_{dark}\left[\left(\frac{P_{light}}{P_{light}^s} + 1\right)^{\frac{\eta}{m}} - 1\right] \quad (6)$$

, in which $I_{dark}$ is the dark current, $m$ is the ratio of the surface potential variation $V_{ph}$ of a channel-continuous nanowire and $\Delta\varphi_{sd}$ of a channel-pinched-off nanowire.[20] Other parameters were defined in previous equations.

Fig.3d exhibits the photoresponsivity of the channel-pinched-off nanowire as a function of light intensity at different gate voltages, which can be analytically described by eq.(7). The term in the curly braces after $R_{max}$ will reach 1 as the light intensity $P_{light}$ decreases to zero. The photoresponsivity can be also plotted a function of gate at different light intensities as shown in Fig.3e. The dependence of photoresponsivity on gate voltage follows a pattern similar to photocurrent in Fig.3b except that the photoresponsivity increases as the light intensity decreases.

The photoresponsivity reaches a maximum $R_{max}$ at zero light intensity which is plotted as the pink line by plugging into $R_{max}$ the gate-dependent dark current $I_{dark}$ in Fig.3a and $P_{light}^S$ found in the inset of Fig.3d.

$$R = \frac{I_{ph}}{P_{light}WL} = R_{max}\left\{\frac{mP_{light}^S}{\eta P_{light}}\left[\left(\frac{P_{light}}{P_{light}^S}+1\right)^{\frac{\eta}{m}}-1\right]\right\} \qquad (7)$$

, in which $R_{max} = \frac{\eta I_{dark}}{mWLP_{light}^S}$ with the parameters defined previously.

The photodetectivity $D^*$ for channel-pinched-off nanowires is written as eq.(8) in which $D^*$ will reach the maximum value $D_{max}^*$ as the light intensity approaches to zero. $D_{max}^*$ is proportional to the minority recombination lifetime $\tau_0$ and the square root of the resistance ($\sqrt{r}$) in darkness. At $V_g$ < ~ 0.8 V, the gate voltage increase $\tau_0$ (the inset of Fig.3c) but decreases $J_{dark}$ (Fig.3a). This competing trend results in a slightly increase of photodetectivity (Fig.3f). At $V_g$ > ~ 0.8 V, $\tau_0$ first decreases and then levels off (the inset of Fig.3c) while the dark current monotonically decreases, resulting in a quick drop followed by a level-off in the photodetectivity (Fig.3f). For this 70nm wide nanowire, $D_{max}^*$ starts from ~ $5 \times 10^{12}$ Jones at low gate voltages and drops to ~ $5 \times 10^{11}$ Jones at high gate voltages.

$$D^* = \frac{R\sqrt{WL}}{\sqrt{\frac{4k_BT}{r}}} = D_{max}^*\left\{\frac{mP_{light}^S}{\eta P_{light}}\left[\left(\frac{P_{light}}{P_{light}^S}+1\right)^{\frac{\eta}{m}}-1\right]\right\} \qquad (8)$$

, in which $D_{max}^* = \frac{2\tau_0\alpha\eta I_{dark}}{m\hbar\omega H n_i}\sqrt{\frac{r}{4k_BTWL}}$.

**Nanowires with a width $2W_{dep}^0 < W < 2W_{dep}^{max}$.** Lastly, we will focus on the third kind of devices that have an intermediate width. It starts with a continuous channel but can be pinched off by gate voltage. Take the 90 nm wide nanowire as an example shown in Fig. 4a. The increase of the applied gate voltage first parabolically reduces the current (see the inset of Fig.4a) because the surface depletion region ($\sim\sqrt{V_g + V_{bi}}$) is widening to narrow down the conduction channel. At a gate voltage higher than ~ 0.8 V, the conduction channel is pinched off and the current drops exponentially (linear in logarithmic scale). Under light illumination, the I-$V_g$ curve right shifts, increasing the threshold voltage. The induced photocurrent, interestingly, first increases to its peak around Vg = 0.4V and quickly drops towards zeros as the gate voltage increases (Fig.4b). The initial increase in photocurrent is caused by mostly the enhancement of electric surface passivation effect, similar to what we observed in Fig.2b for the channel-continuous nanowires. However, before the surface depletion region reaches its theoretical maximum, the conduction channel is pinched-off and a potential barrier is created between source and drain. As a result, the photocurrent starts to drop, consistent with the behavior of the channel-pinched-off nanowires in Fig. 3b.

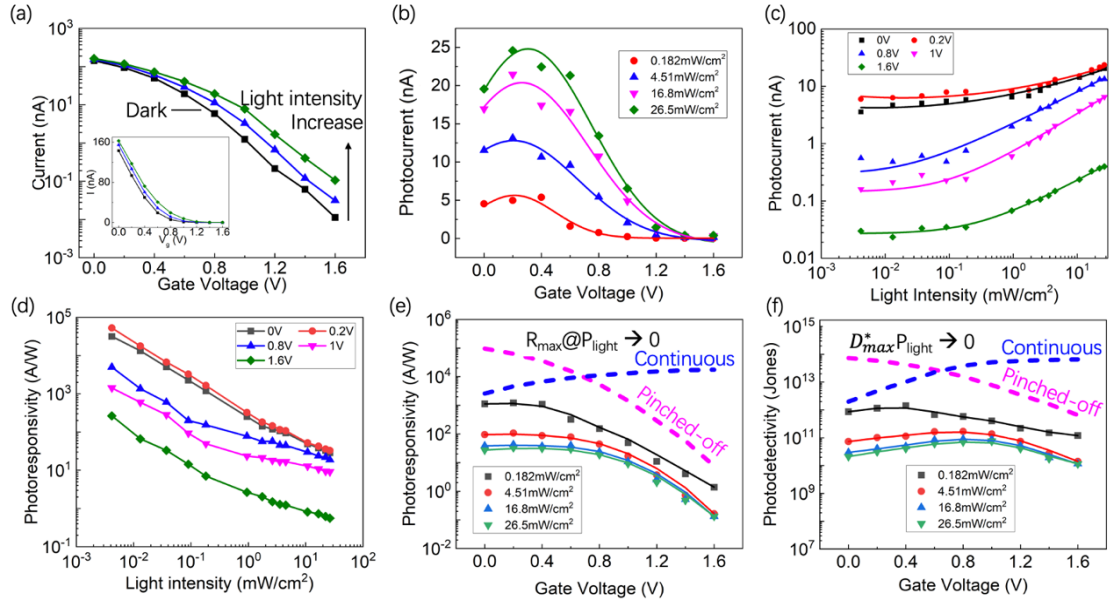

Figure 4. Gate transfer optoelectronic characteristics for the nanowire device 90 nm wide at a fixed bias of 1V. (a) Current *vs* gate voltage under different light intensity. (b) Photocurrent *vs* gate voltage under different light intensity. (c) Photocurrent *vs* light intensity under different gate voltage. (d) Photoresponsivity *vs* light intensity under different gate voltage. (e) Photoresponsivity *vs* gate voltage under different light intensity. The dash line represents the corresponding limit for channel-continuous and pinch-off mode when light intensity approaches zero. (f) Photodetectivity vs gate voltage under different light intensity. The dash line represents the corresponding limit for channel-continuous and pinch-off mode when light intensity approaches zero.

Fig.4c plots the photocurrent as a function of light intensity which can be well fitted with eq.(2) at small gate voltages for channel-continuous nanowires and eq.(6) at large gate voltages for pinched-off nanowires. The photocurrents in between are difficult to nicely fit with either eq.(2) or eq.(6). Similar phenomena were also observed for photoresponsivity dependent on light intensity in Fig.4d.

Photoresponsivity and photodetectivity dependent on gate voltage are plotted in Fig.4e and f, respectively. At small gate voltages, these parameters increase as the gate voltage, largely following the behavior of a channel-continuous nanowire (Fig.2e and f), in particular at high light intensity. At large gate voltages, these parameters decrease as the gate voltage increases, similar to what we observed for the channel-pinched-off nanowire in Fig.3e and f. Interestingly, these parameters will reach a maximum at a medium voltage, for example ~ 0.8 V, in particular under high light intensity.

Conclusion

In this work, we investigated the gate modulated photoresponses of silicon nanowire photoconductors. Depending on whether the channel is continuous or pinched off, the gate voltage had a different impact on the photoresponses of the nanowires. For the nanowires that are too wide to pinch off, the gate voltage enhances the photoresponses by strengthening the electric passivation effect. The enhanced photoresponses reach a plateau as the surface depletion region enters the strong inversion mode at high gate voltage. For the narrow nanowires that are pinched off by the small width, the gate voltage will not enhance the electric passivation effect but increase the potential barrier between source and drain, resulting in smaller photoresponses at higher gate voltage. For those with an intermediate size, the nanowires transit from a continuous channel to a pinched-off one with the photoresponses reaching a peak during this transition.

From these observations, we conclude that a nanowire photoconductor will have the best performance when the nanowire surfaces are in strong inversion and the nanowire channel is just pinched off by its physical width. A strongly inverted surface will maximize the electric passivation effect and hence the photoresponse. When the channel is just pinched off, the dark current is suppressed without sacrifice of photoresponse, resulting in a maximum of photodetectivity and photoresponsivity.

**Experimental**

**Device fabrication.** The silicon-on-insulator (SOI) wafers were first cleaned with acetone and deionized (DI) water. Boron ions were then implanted into the device layer of the SOI wafer at an implantation energy of 30 keV and a dose of $2.2 \times 10^{13}$ cm$^{-2}$. The doping peak is located at 110 nm below the surface with a maximum concentration of $8.7 \times 10^{17}$ cm$^{-2}$. After the implantation, rapid thermal annealing (RTA) was employed to activate the baron atoms at 1000°C for 20s. Then the silicon wafers were cut into 2×2 cm$^2$ squares. All pieces were cleaned with piranha solution (98% $H_2SO_4$:30% $H_2O_2$, 3:1(v/v). A 250 nm-thick layer of poly(methyl methacrylate)(PMMA) resist (XR-1541-006, Dow Corning Electronics, USA) was spin-coated on the p-type SOI samples at 4000 rpm for 60s. Then the wafers were baked at 180°C for 90s. The PMMA resist was exposed by electron beam lithography (Vistec EPBG5200) and subsequently developed in MIBK and IPA. PMMA is a positive electron beam resist. After that, aluminum (Al) was evaporated to the exposed region to form the etch mask for micropads. Then, NR9-1500PY (Futurrex Inc. USA) photoresist was coated on the wafers at 4000 rpm for 40s. After baked at 140°C for 60s, the NR9 resist was exposed to UV light (MDA-400) and developed in the RD6 developer after post-baking at 110°C for 60s. After that, Al film was evaporated, followed by a liftoff process. The Al nanowire and micropad patterns defined by electron beam exposure and photolithography were then transferred to the SOI device layer by reactive ion etching (RIE, Sentech ICP Reactive Ion Etching System), forming an array of silicon nanowires with each connecting to two silicon pads for metal contacts. After photolithography, 20nm thick cobalt and 180nm thick aluminum were evaporated onto the silicon pads as electrodes. To form good ohmic contacts, the wafers were then annealed in argon atmosphere for 15min at 230°C. Then a 50nm thick layer of $HfO_2$ was then grown of the surface of the sample at 250°C by plasma-enhanced atomic layer deposition (PEALD, Beneq). Another photolithography was performed to expose the electrodes parts. Inductively coupled plasma (ICP) Reactive Ion Etching (RIE) System was applied to etch the $HfO_2$ to expose the electrodes. The third photolithography was conducted to create the liftoff pattern for the ITO gate electrode, followed by the ITO evaporation and lift-off. In the end, the device samples were annealed in argon atomsphere for 10mins at 400°C to increase the transparency and conductivity of ITO.

**Photoresponse measurements.** The devices were characterized in a probe station by high-precision digital sourcemeters (Keithley 2636). Light from a commercial LED (λ = 532 nm) created a uniform light illumination spot (~5 mm in diameter) on the samples. The light intensity was controlled by the driving current from a sourcemeter (Keithley 2400). The light intensity was calibrated by a commercial photodiode (G10899-003 K, Hamamatsu).

Acknowledgement

This work was financially supported by the special-key project of Innovation Program of Shanghai Municipal Education Commission (No. 2019-07-00-02-E00075), the National Science Foundation of China (NSFC), (No. 92065103), the Oceanic Interdisciplinary Program of Shanghai Jiao Tong University (No. SL2022ZD107), the Shanghai Jiao Tong University Scientific and Technological Innovation Funds (No. 2020QY05), and the Shanghai Pujiang Program (No. 22PJ1408200). The devices were fabricated at the Ccenter for Advanced Electronic Materials and Devices (AEMD), Shanghai Jiao Tong University.

AUTHOR DECLARATIONS

The authors have no conflicts to disclose.

DATA AVAILABILITY

The data that support the findings of this study are available from the corresponding authors upon reasonable request.